\documentclass{article} 
\usepackage{iclr2024_conference,times}

\usepackage{amsmath,amsfonts,bm}

\def\eqref#1{equation~\ref{#1}}

\def\1{\bm{1}}

\DeclareMathAlphabet{\mathsfit}{\encodingdefault}{\sfdefault}{m}{sl}
\SetMathAlphabet{\mathsfit}{bold}{\encodingdefault}{\sfdefault}{bx}{n}

\usepackage{hyperref}
\usepackage{url}
\usepackage{graphicx}

\usepackage{caption}
\usepackage{subcaption}

\usepackage{lipsum}
\usepackage{xcolor}

\usepackage{comment}

\title{Diffusion-Based Joint Temperature and Precipitation Emulation of Earth System Models}

\author{Katie Christensen$^1$, Lyric Otto$^1$, Seth Bassetti$^2$, Claudia Tebaldi$^3$, Brian Hutchinson$^{1,4}$ 
\\
$^1$Department of Computer Science, Western Washington University, Bellingham, WA\\
$^2$ Computer Science Department, Utah State University, Logan, UT \\
$^3$Joint Global Change Research Institute, Pacific Northwest National Laboratory, College Park, MD \\
$^4$AI and Data Analytics Division, Pacific Northwest National Laboratory, Richland, WA\\
\texttt{\{chris90, ottol, brian.hutchinson\}@wwu.edu,}\\
\texttt{seth.bassetti@usu.edu, claudia.tebaldi@pnnl.gov}}

\iclrfinalcopy 
\begin{document}

\maketitle

\begin{abstract} 
Earth system models (ESMs) are the principal tools used in climate science to generate future climate projections under various atmospheric emissions scenarios on a global or regional scale. Generative deep learning approaches are suitable for emulating these tools due to their computational efficiency and ability, once trained,  to generate realizations in a fraction of the time required by ESMs. We extend previous work that used a generative probabilistic diffusion model to emulate ESMs by targeting the joint emulation of multiple variables, temperature and precipitation, by a single diffusion model. 
    Joint generation of multiple variables is critical to generate realistic samples of phenomena resulting from the interplay of multiple variables.  
    The diffusion model emulator takes in the monthly mean-maps of temperature and precipitation and produces the daily values of each of these variables that exhibit statistical properties similar to those generated by ESMs. Our results show the outputs from our extended model closely resemble those from ESMs on various climate metrics including dry spells and hot streaks, and that the joint distribution of temperature and precipitation in our sample closely matches those of ESMs.
\end{abstract}

\section{Introduction}
Earth system models (ESMs) simulate large scale phenomena and extreme weather events and provide insights into the effects of human activity on Earth's climate at the global and regional scales. These models can offer decision-makers crucial insights for addressing the impacts of future climate scenarios on various systems, including energy and land use. Due to the rarity of extreme events, ESMs must be run numerous times to obtain enough realizations for robust statistics of rare events. This poses an issue when considering ESMs' significant computational demand and multiple experimental uses. Emulators address this issue by producing realistic future climate projections across various emissions scenarios while significantly reducing the time and computational resources needed, since, once trained, they can generate many realizations in a computational efficient way \citep{Kasim21Building}. Generative deep learning approaches, in particular, have emerged as strong candidates for developing such emulators \citep{Addison22Machine, Saharia22Photorealistic, Ho22Video}.

We extend the DiffESM model \citep{Bassetti23DiffESM, Bassetti23code}, which emulates ESMs with generative, probabilistic diffusion models. We integrate multiple variables, specifically temperature and precipitation, into the DiffESM pipeline \citep{Bassetti_2024} to model the variables' joint spatio-temporal trends. Our model generates realizations of future behavior that mimics ESM output, and from which statistically robust estimates of metrics related to extreme weather events can be efficiently computed. We apply our emulator to one ESM under a scenario of future greenhouse gas emissions. In the original, univariate version, DiffESM produces month-long samples of either the daily mean temperature or daily total precipitation. The monthly mean maps used for its conditioning can be generated from other low-cost emulators, such as STITCHES \citep{Tebaldi22Stitches} or MESMER \citep{Nath22Mesmer}; DiffESM complements  these existing emulator approaches. The emulation of single variables, however, may fail to represent the coherent relationship between variables, particularly important for the severity of extreme events that result from the combination of multiple factors, like droughts made more severe by high temperatures, or vice-versa. In our extension, DiffESM is trained to emulate these two variables, producing realistic month-long samples of both daily precipitation and daily average temperatures that capture their covariance and are consistent with their monthly means. 

Additional applications of machine learning techniques in the atmospheric and climate sciences include weather forecasting \citep{Wang19Deep, Scher18Predicting, Rasel18An} and downscaling, i.e., increasing the spatial resolution of ESM output by borrowing information from regional models or observations on a finer scale \citep{Hobeichi23Using, Hassen21Novel, Jebeile21Understanding}. Generative adversarial networks (GANs) have shown their value in this field of application \citep{Puchko20Deepclimgan, Ayala21Loosely, Kashinath21Physics}. For example, \cite{Hess2022Physically} uses GANs to improve the spatial resolution of precipitation generation in finer grained locations that were previously susceptible to bias in ESMs. However, GANs have proven to be relatively difficult to train, and the easier-to-train diffusion models have become a more popular choice for generative tasks \citep{Dhariwal21Diffusion}. 
While the statistical community has long addressed the problem of modeling the joint distribution of climate variables, their efforts have predominately focused on more limited scales, both temporally and spatially, for example, through the use of copula models \citep{Bevacqua17Multivariate, Li22Development, Sarhadi18Multidimensional}.

\section{Methods}
\subsection{DiffESM Background} \label{sec:diffesm_background}
The original DiffESM in \cite{Bassetti23DiffESM, Bassetti_2024} is a denoising generative probabilistic model. It generates samples through iterative denoising steps from a known Gaussian distribution to the target's unknown distribution~\citep{Ho20Denoising}. The model architecture of DiffESM is influenced by Video Diffusion \citep{Ho22Video} and Imagen Video \citep{Ho22Imagen}. It contains a fully convolutional U-Net \citep{Ronneberger15UNet} for each denoising step, and temporal and spatial convolution layers. The input shape of the denoising step is $T \times H \times W$, whose elements contain the variable of interest (modelled individually). $T = 28$ is a 28-day sequence length treated as a ``month," and $H=96$ and $W=96$ are the number of grid-boxes in the longitude and latitude dimension, i.e., the ESM grid resolution (corresponding to about $350\mbox{ km}$ in the longitude dimension at the equator, and half that length in the latitude dimension). Each denoising step maintains the same shape as its training samples, i.e., consists of $96 \times 96$ grids of daily temperature (or precipitation) along a 28-day coherent temporal sequence. The dataset used to train DiffESM consist of daily mean temperatures (Celsius) and daily precipitation (mm) from the Institut Pierre-Simon Laplace Earth System Model (IPSL-CM5A) \citep{Dufresne13Climate}. DiffESM was also trained and validated, separately for the homologous output from the Community Earth System Model (CESM1-CAM5) \citep{Kay15The} whose resolution is higher ($256 \mbox{ by } 128$, corresponding to just over $100\mbox{ km}$). To prepare the data, the temperature units are converted to degrees Celsius and normalized by dividing all values by twenty, which suffices to mitigate numerical instability. Additionally, logarithmic normalization ($\log(x+1)$ is applied to the precipitation values, managing extremes. 

\begin{figure}
  \centering
  \includegraphics[width=0.6\linewidth]{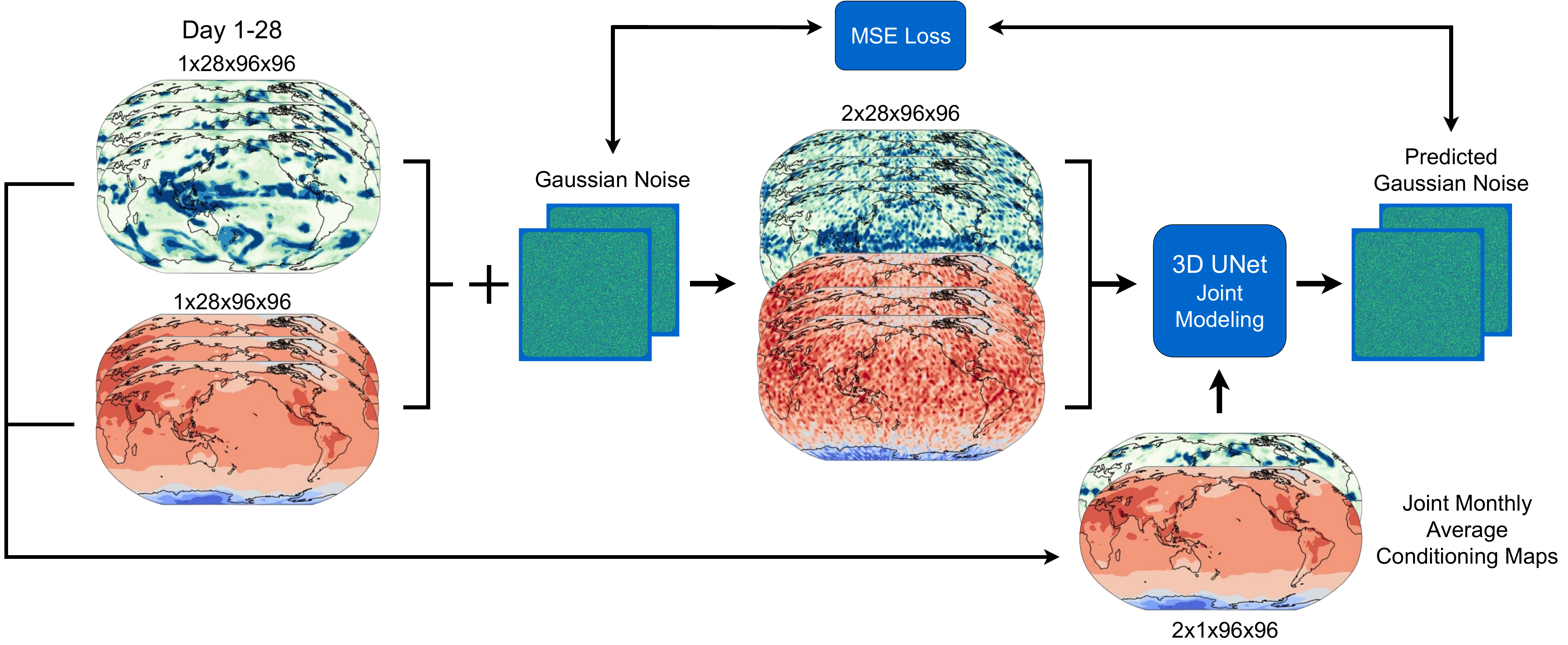}
  \caption{Training loop, illustrating the input and output channel $C=2$ for both daily temperature and precipitation.}
  \label{fig:trainingloop}
\end{figure}

\subsection{Multiple Variable Modeling}
We extend the work of DiffESM by integrating multiple variables into the generative process (both in the conditioning and generated samples). In this extension, the samples being denoised are $C \times T \times H \times W$, where $C=2$ is a channel dimension, containing both daily temperature and precipitation variables. Figure~\ref{fig:trainingloop} shows the training loop of our model, where the input consists of both the daily precipitation and average temperature sequence, and where the model outputs the generated 28-day month-long sequences for each variable. Once trained, the diffusion model takes as input two consistent monthly mean maps of temperature and precipitation and generates two 28-day sequences of the variables that will (a) maintain joint, spatial and temporal coherence, learned from the ESM behavior, and (b) be consistent with the input maps. In order to facilitate comparison to prior work, we train and evaluate this extended work on the IPSL-CM5A ESM run under the RCP scenarios of CMIP5 \citep{Mossetal2010}, with intentions to move to CMIP6 \citep{ONeilletal2016} for future developments. The $T, H,$ and $W$ are the same as reported in \cite{Bassetti23DiffESM, Bassetti_2024}, and our data normalization (see Sec.~\ref{sec:diffesm_background}) is the same, as well.
We use a learning rate of 0.0001, a batch size of 64 across four GPUs, and we train for 10 epochs using the Adam optimizer \citep{Kingma15Adam}. 

\section{Experiments}
Our dataset consists of both daily precipitation and daily mean temperature outputs from the IPSL-CM5A ESM on a $96 \times 96$ grid. We use a total of 6 ESM realizations per dataset, each of them consisting of both temperature and precipitation daily output. The ESM experiments cover the period from pre-industrial times to 2100, using ``historical'' anthropogenic forcings from 1850 to 2006, and the RCP8.5 scenario (the most extreme emissions scenario) from 2006 to 2100. We use four of the six bivariate realizations as the training set, one realization as the first held out (Held Out 1) realization, and the remaining one realization as the second held out (Held Out 2) realization (see below for the usage of the latter two). 

\subsection{Metrics}
After training, we use the Held Out 1 realization to create pairs of monthly mean maps by averaging temperature and precipitation over each 28-day ``month'' in the range 2080-2100. We feed each pair of monthly mean maps to the diffusion model to generate one 28-day bivariate sample of daily temperature and precipitation per month. We rely upon these monthly means to preserve the temporal coherence from month to month, while the diffusion model learns to preserve the intra-month coherence of the daily timeseries. We repeat this process using the Held Out 2 realization and generate an additional 28-day bivariate sample per month. In the absence of specification, reference to a ``generated sample'' pertains to a sample conditioned on Held Out 1 monthly means. 

We compute temperature and precipitation metrics (summaries of daily behavior within each sequence of 28 days) per sample for each spatial location, then average over the months to produce metric maps. Likewise, we produce metric maps for the Held Out 1 and Held Out 2 sets during the same time range. We evaluate the performance of the emulator by comparing these metric maps; specifically, we take the difference of the generated samples and Held Out 2, and compare that to the difference between Held Out 1 and Held Out 2. The latter, comparing two realizations from the ESM that are different only because of internal noise gives us a measure of the unavoidable difference between realizations. To summarize the discrepancies between the generated or Held Out 1 samples and the Held Out 2 samples, we also produce superimposed histograms of the differences at all the spatial locations in these difference maps. We repeat this process using the generated samples from the Held Out 2 monthly means.

For three individual locations, representative of different climates (the ESM grid points closest to Honolulu, Hawaii; Melbourne, Australia; and Novosibirsk, Russia) we also produce contour maps to show the correlation between the two variables using the generated sample and compare to the samples from the Held Out 1 and Held Out 2 realizations. We separate the samples into dry and wet days where precipitation is $<1.00$ mm and $>=1.00$ mm, respectively. We then calculate the deciles of the precipitation and temperature data from the Held Out 2 sample and use these to partition the days where the temperature and precipitation fall within each decile, and calculate the fraction of days within each partition relative to the total number of wet days in the map. We plot these empirical joint distributions as contour maps and heatmaps to show the difference between the variable correlation of the generated sample compared to the correlations found in Held Out 1 and Held Out 2. We also produce histograms of the temperature distribution of dry days over the same temperature deciles from the wet days of Held Out 2. 

\begin{figure}
  \centering
  \includegraphics[width=0.9\linewidth]{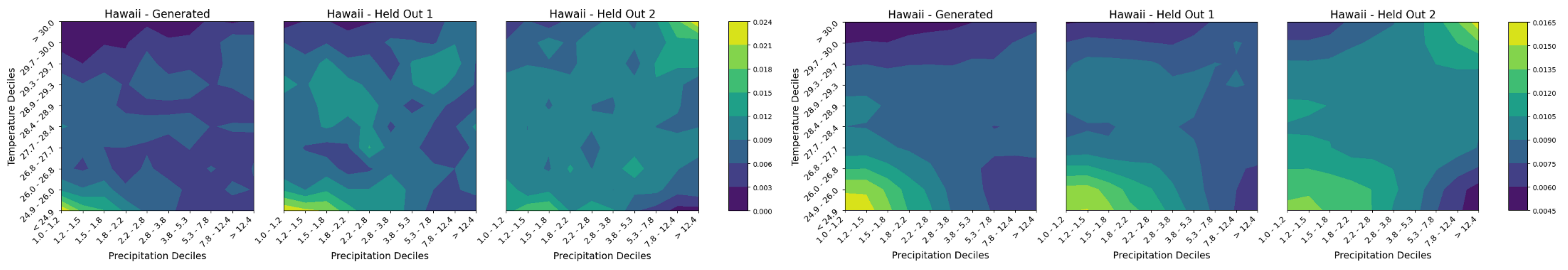}
  \caption{Joint variables discretized into temperature and precipitation deciles computed from the Held Out 2 realization. The distribution is computed only over wet days (when precipitation $>=1.00$ mm), for location Hawaii. The sets of contour maps contain no smoothing (left) and average smoothing with a $3 \mbox{ by } 3$ kernel (right). Each set compares the distributions of the generated (left), Held Out 1 (middle), and Held Out 2 (right). Distributions are computed from 252 28-day samples.}
  \label{fig:hawaii_contour}
\end{figure}
\begin{figure}
  \centering
  \includegraphics[width=0.8\linewidth]{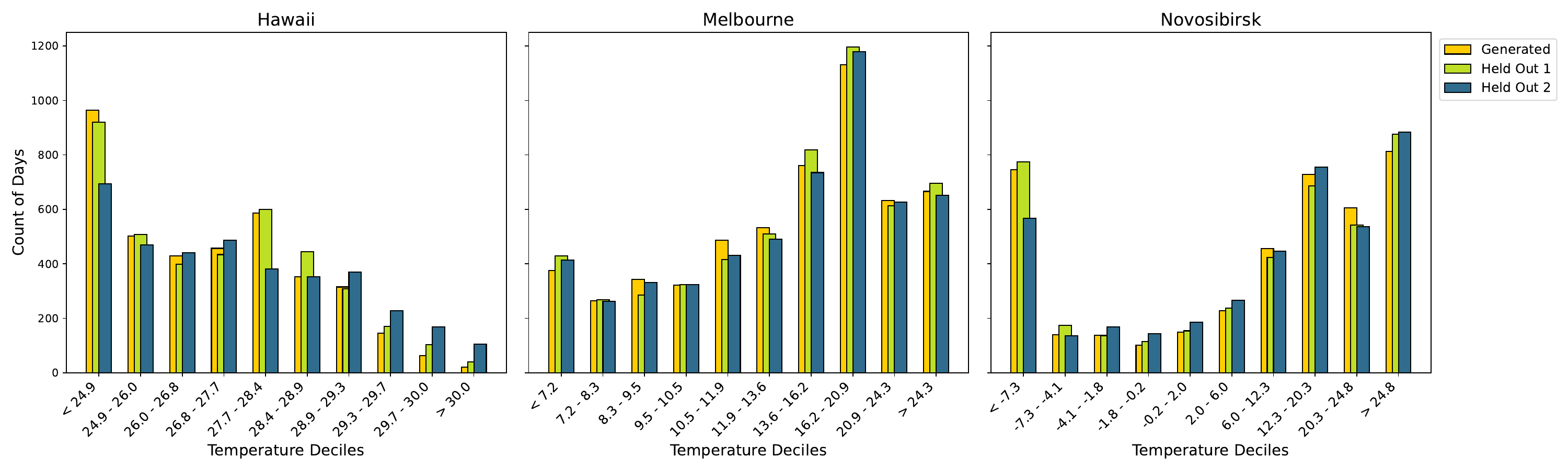}
  \caption{Joint distribution of variables discretized into temperature deciles computed from the Held Out 2 realization. The distribution is computed only over dry days (when precipitation $<1.00$ mm), for locations Hawaii (left), Melbourne (middle), Novosibirsk (right). The three bars of the figure compare the distributions of the generated realizations (yellow) to that of the Held Out 1 (green) and Held Out 2 (blue). Distributions are computed on the basis of 252 28-day samples.}
  \label{fig:histograms}
\end{figure}

\subsection{Results}
Our results demonstrate the coherent relationship between variables from the generated sample of the IPSL-CM5A dataset. Figure~\ref{fig:hawaii_contour} reveals these trends across wet days where precipitation is $>=1.00$ mm. The generated sample is shown on the left, Held Out 1 in the middle, and Held Out 2 on the right for the location Hawaii. Appendix \ref{sec:Additional} shows the additional locations Melbourne (top) and Novosibirsk (bottom) in Figure \ref{fig:melbourne_novosibirsk_contour}, as well as the heatmaps across all three locations in Figure \ref{fig:all_locations_heatmaps}. The contour maps exhibit similar peak and valley patterns between the generated, Held Out 1, and Held Out 2, reflecting agreement in the bivariate distribution. Figure \ref{fig:histograms} shows the distribution of dry days where precipitation is $<1.00$ mm. It is noticeable that the model marginally overpredicts temperature on cool, dry days and slightly underpredicts temperature on warm, dry days. It is apparent that these variables are interdependent as we observe the correlation between high temperatures and low precipitation, as well as low temperatures and high precipitation. 

\begin{figure}
  \centering
  \includegraphics[width=\linewidth]{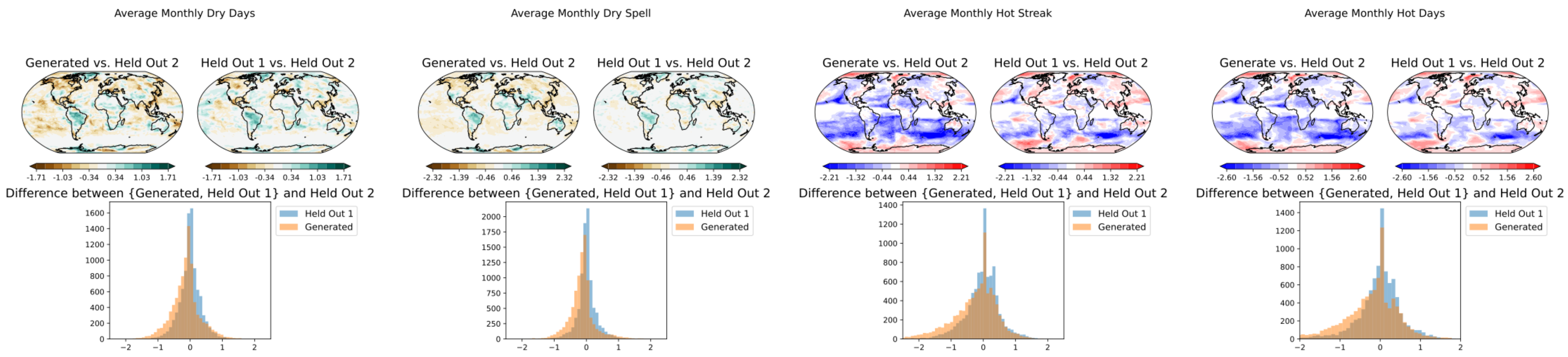}
  \caption{Pairs of difference maps between generated and Held Out 2 (left) and Held Out 1 and Held Out 2 realizations (right), and the corresponding superimposed grid-box error histograms of generated and Held Out 2 (orange) and Held Out 1 and Held Out 2 (blue) for two precipitation metrics (left) and two temperature metrics (right) for the bivariate sample conditioned on the Held Out 1 monthly means.}
  \label{fig:diffmaps-bivariate}
\end{figure}

We also plot four pairs of difference maps in Figure~\ref{fig:diffmaps-bivariate} for the precipitation metrics average monthly dry days and dry spell, and the temperature metrics average monthly hot streak and hot days. For each pair, the generated minus Held Out 2 is shown on the left and the Held Out 1 minus Held Out 2 is on the right. We also compare the difference distributions and find these to be similar between the generated minus Held Out 2 (orange) and the Held Out 1 minus Held Out 2 (blue). We plot these metrics in Figure~\ref{fig:diffmaps-bivariate} for the bivariate generated sample conditioned on the Held Out 1 monthly means. We do expect there to be some variability between these, and our results show a similar magnitude of variability between the generated sample and Held Out 2 to the naturally occurring variability between Held Out 1 and Held Out 2. These trends are reflected across the climate metrics hot streak and dry spell, demonstrating the model's capability of capturing the intra-month temporal coherence of ESMs. Appendix \ref{sec:Additional-DiffMaps} discusses additional analyses, comparing to the univariate generated sample and the bivariate generated sample conditioned on Held Out 2.
\section{Conclusions and Future Work}
We show that integrating joint temperature and precipitation emulation into the DiffESM generative diffusion model effectively produces joint temperature-precipitation samples that not only match the ESM in the marginal distributions, but also in their joint distribution. Specifically, we observe results similar to those reported in previous work \cite{Bassetti23DiffESM, Bassetti_2024} on single variable generation for various climate metrics based on daily behavior, including dry spells and hot days. We further demonstrate the ability to recreate the interrelationship between temperature and precipitation. This is a promising result as there are clear advantages in incorporating multiple variables into the ESM emulator and producing consistent joint realizations; this represents a step forward in the generation of realistic samples of climate scenarios. The joint behavior of variables is particularly important for addressing the emulation of extreme events that depend on both variables exacerbating hazardous conditions, like heat extremes made more dangerous by humidity, droughts made more severe by heat, and vice-versa heat domes made more persistent by dryness of the land surface.  
Along these lines, further experimentation includes integrating additional variables such as daily relative humidity, and daily high and low temperatures. We also plan to train and evaluate on additional ESMs including the CESM1-CAM5 ESM and the MIROC ESM \cite{Watanabe11MIROC}. Finally, we are working on emulating multiple ESMs with a single diffusion model, capable of producing samples in the style of any given constituent model, but potentially benefiting from having been trained across a larger set of ESMs, realizations,  and scenarios.

\subsubsection*{Acknowledgments}
This research was supported by the U.S. Department of Energy, Office of Science, under the MultiSector Dynamics, Earth and Environmental System Modeling Program. The Pacific Northwest National Laboratory, operated by Battelle Memorial Institute under contract DE-AC05-76RL01830, facilitated this research. The authors express gratitude to the World Climate Research Programme's Working Group on Coupled Modelling for CMIP and extend appreciation to the climate modeling team at Institut Pierre Simon Laplace (France) for sharing its model outputs. Coordinating support for CMIP was provided by the U.S. Department of Energy's Program for Climate Model Diagnosis and Intercomparison, in collaboration with the Global Organization for Earth System Science Portals, which also contributed to the development of software infrastructure.

\newpage
\bibliography{iclr2024_conference}
\bibliographystyle{iclr2024_conference}

\appendix
\newpage
\section{Additional Contour Plots and Heatmaps}
\label{sec:Additional}

\begin{figure}[h]
  \centering
  \includegraphics[width=0.9\linewidth]{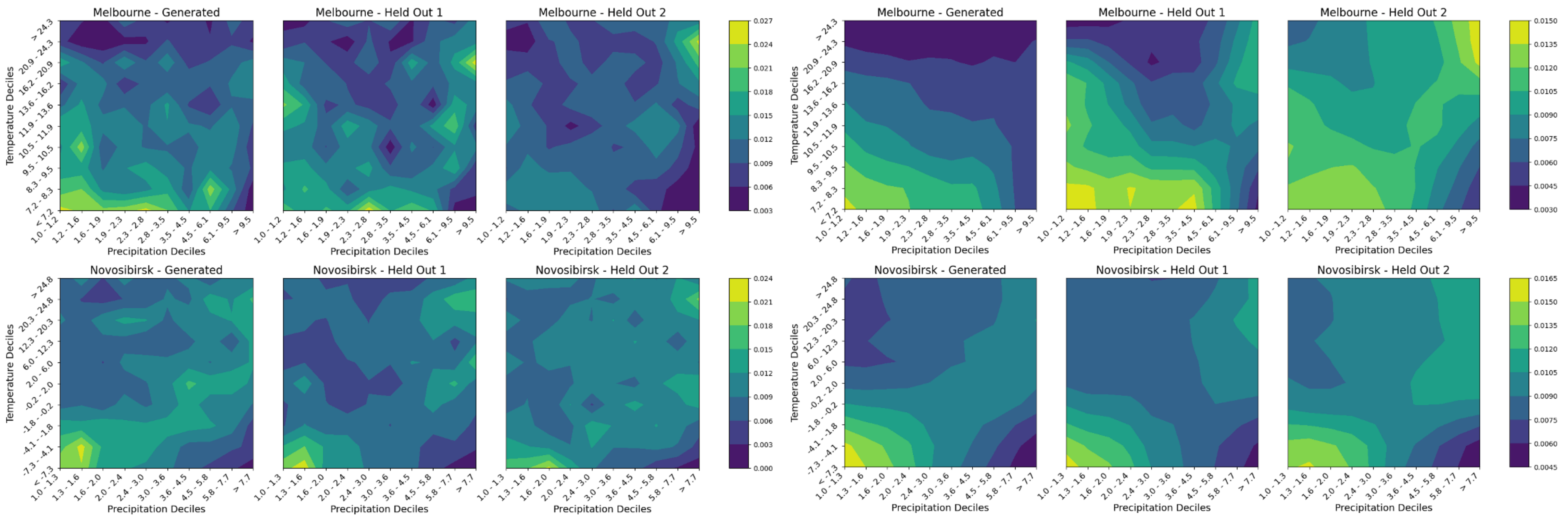}
  \caption{Joint distribution of temperature and precipitation discretized into temperature and precipitation deciles computed from the Held Out 2 realization over wet days (when precipitation $>=1.00$ mm), for locations Melbourne (top), Novosibirsk (bottom). The set of contour maps on the left contain no smoothing, the set on the right contain average smoothing with a $3 \mbox{ by } 3$ kernel. The three columns of the figure compare the distributions of the generated realizations (left) to that of Held Out 1 (middle) and Held Out 2 (right). Distributions are computed on the basis of 252 28-day samples.}
  \label{fig:melbourne_novosibirsk_contour}
\end{figure}

\begin{figure}[h]
  \centering
  \includegraphics[width=0.7\linewidth]{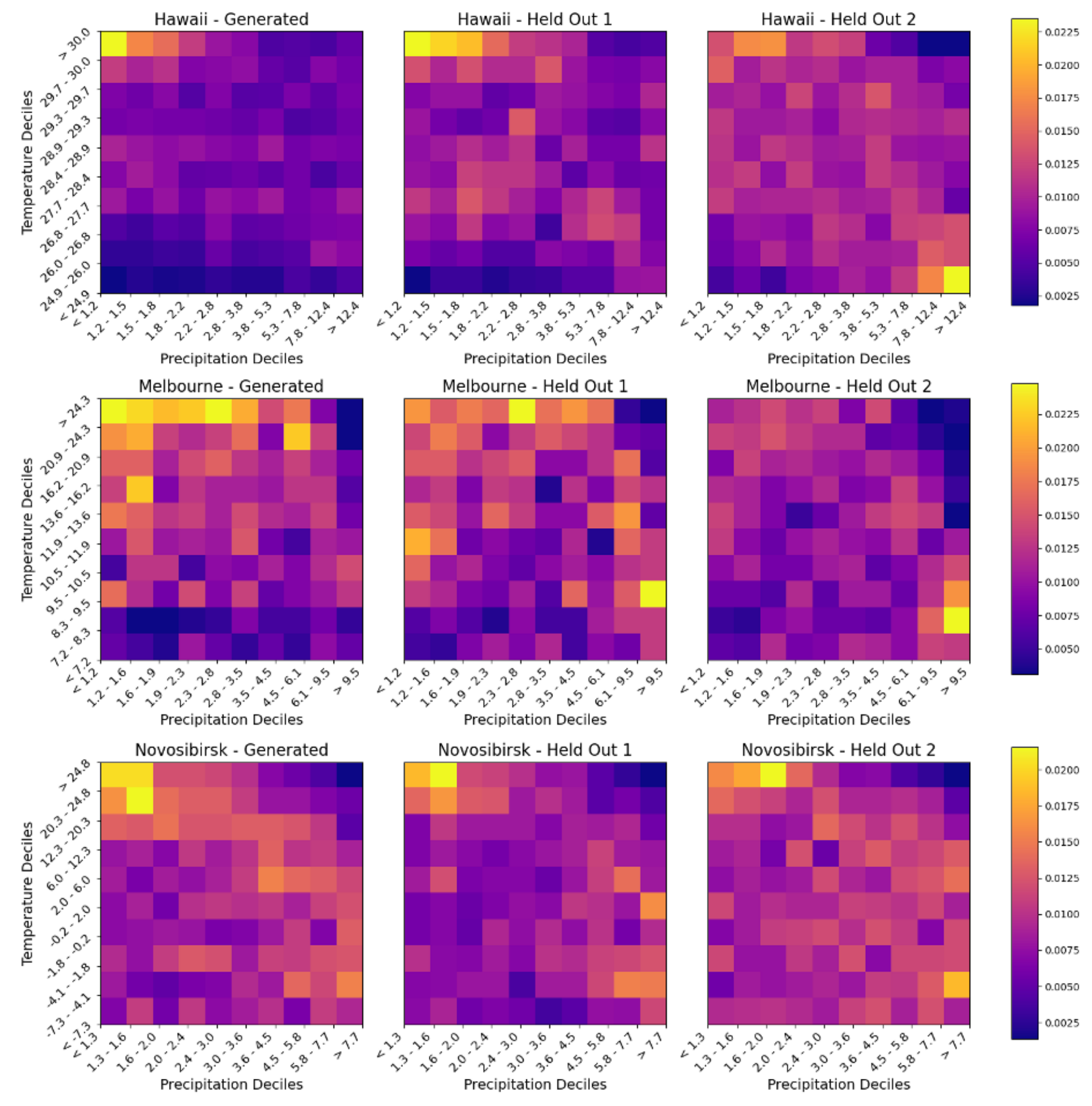}
  \caption{Joint distribution of temperature and precipitation discretized into temperature and precipitation deciles computed from the Held Out 2 realization over wet days (when precipitation $>=1.00$ mm), for locations Hawaii (top), Melbourne (middle), Novosibirsk (bottom). The three columns of the figure compare the distributions of the generated realizations (left) to that of Held Out 1 (middle) and Held Out 2 (right). Distributions are computed on the basis of 252 28-day samples.}
  \label{fig:all_locations_heatmaps}
\end{figure}

\newpage
\section{Additional Difference Maps}
\begin{figure}[h]
  \centering
  \includegraphics[width=\linewidth]{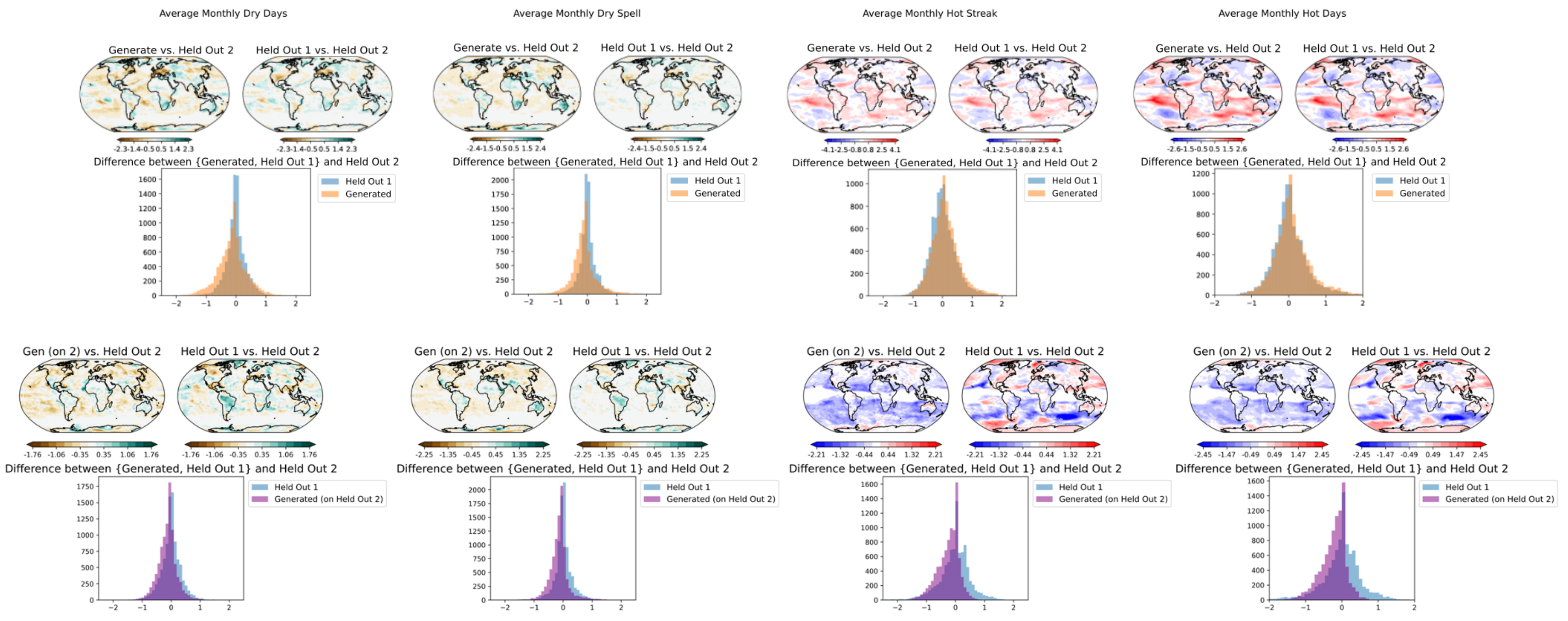}
  \caption{Pairs of difference maps between generated and Held Out 2 (left) and Held Out 1 and Held Out 2 realizations (right), and the corresponding superimposed grid-box error histograms of generated on Held Out 1 and Held Out 2 (orange), or generated on Held Out 2 and Held Out 2 (purple), and Held Out 1 and Held Out 2 (blue), for two precipitation metrics (left) and two temperature metrics (right) for the single variable samples (top), and the generated samples conditioned on the Held Out 2 monthly means (bottom).}
  \label{fig:diffmaps_supmaterials}
\end{figure}

Figure~\ref{fig:diffmaps_supmaterials} replicates the metric map experiment using a univariate DiffESM model (top) and the bivariate model, but conditioned on Held Out 2 (bottom). For the bottom row, the histograms compare the difference distributions of the generated on Held Out 2 minus Held Out 2 (purple), and the Held Out 1 minus Held Out 2 (blue). We expect the generated on Held Out 2 minus Held Out 2 to exhibit a narrower distribution than Held Out 1 minus Held Out 2 due to the variability between the two held out realizations, while still maintaining a degree of variability owing to the randomness of the generative process. These expectations are consistent with the results in Figure~\ref{fig:diffmaps_supmaterials}.

\label{sec:Additional-DiffMaps}

\end{document}